# Edge Truncation Effect Suppression of Ultrawideband Phased Arrays for Radar Application

Chenglong Fan, Shi-Wei Qu, *Senior Member, IEEE*, Shiwen Yang, *Fellow, IEEE*, and Jun Hu, *Senior Member, IEEE*

*Abstract*—This paper presents a novel effective method to suppress the edge truncation effect of ultrawideband tightly coupled dipole linear arrays. To restrain the edge truncation effect within an ultrawideband operating band, a new type of T-shaped metal strip with a resistor is further loaded on the array edges apart from extending the length of the overlapping patches. Besides, the excitation phase of the elements at the array edges is optimized. Full-wave simulation results show that the active standing wave standing ratio of the 2×16 tightly coupled dipole linear arrays using the proposed method is significantly optimized to less than 3.5 within a 5:1 (1.2~6 GHz) bandwidth, while scanning up to ±60° in the E-plane. The effectiveness of the proposed method is experimentally verified by a 2×16 linear array prototype.

## I. INTRODUCTION

BROADBAND phased array antennas, as the front end of radar systems, are essential for enhancing radar resolution, target classification and recognition ability, anti-jamming performance and reliability [1], [2], [3]. Tightly coupled dipole phased arrays (TCDAs), as a representative of broadband phased array antenna, are playing an increasingly important role in radar systems due to their merits of ultrawideband operating band, large scanning angles and low profile. Over the past decade, various types of TCDAs have been proposed and studied [4-6]. To accelerate the solution speed and save computing resources, the design method of TCDAs is mainly based on antenna element simulations in periodic environments, which is equivalent to an infinite array. However, for a finite array, the environment of the edge elements is significantly different from that of the elements in the infinite array, which will inevitably lead to a sharp deterioration of the impedance matching. When the array is used for power transmitting propose, the impedance matching deterioration of the edge elements will result in a large amount of energy reflection, which is fatal for the feeding T/R components [7]. Therefore, it is necessary to find an effective method to suppress the edge truncation effect (ETE) of ultrawideband finite-size TCDAs.

To suppress the ETE of TCDAs, many effective methods have been proposed over the past several years. Significantly increasing the number of the non-excited elements in the array peripheries is a straight-forward way to restrain the ETE [8], but it is unpractical due to a huge waste of the physical array aperture. In [9], the ETE is mitigated by optimizing the antenna element design for a larger bandwidth than required, but this approach increases the complexity and the cost of the element. Ref [10] restrains the ETE by adding shorted elements (extending dipole arms) on the peripheries of the array, achieving an H-plane scan of 30°, but the operating bandwidth is limited to 3:1 (200~600 MHz). In [11], the excitation amplitude and phase of each array element are optimized to suppress the ETE, achieving a 10:1 operating bandwidth while scanning up to ±45° with an active voltage standing wave ratio (VSWR) less than 3. Besides, there are some studies [12], [13] to improve the active VSWR of the finite arrays by using amplitude taper over the array aperture. However, the method based on amplitude optimization will inevitably reduce the maximum power capacity and aperture efficiency of the array. Therefore, it is vital to explore a low-cost, simple and feasible way to restrain the ETE within an ultrawideband operating band and large scanning angles.

In this letter, a novel effective method to restrain the ETE of finite-size tightly coupled dipole linear arrays (TCDLAs) is proposed. Based on extending the length of the overlapping patches, a new type of T-shaped metal strip with a resistor, which is simple and ease of fabrication, is loaded on the array edges. Additionally, the excitation phase of the elements at the array edges is optimized. The TCDLA with the proposed method can feature an operating frequency bandwidth of 5:1 (1.2~6 GHz) with an active VSWR<3.5 while scanning up to ±60° in the E-plane. The above results are experimentally verified using a 2×16 TCDLA prototype.

## II. ARRAY DESIGN

### A. Array Structure

The geometries of the proposed antenna array are shown in Fig. 1, consisting of 16 antenna subarrays placed along the *y*-axis and edge loading structures on both sides. Each antenna subarray, which is designed based on [14], is composed of two antenna elements with equal amplitude and phase feeding, as well as two wide-angle impedance matching (WAIM) layers. Each antenna element consists of a Marchand balun to achieve impedance transformation and unbalanced-to-balanced

This work was supported by the National Natural Science Foundation of China Projects under Grant U20A20165. (Corresponding author: Shi-Wei Qu)

The authors are with the School of Electrical Science and Engineering, University of Electronic Science and Technology of China, Chengdu 611731, China. (email: shiweiqu@uestc.edu.cn).

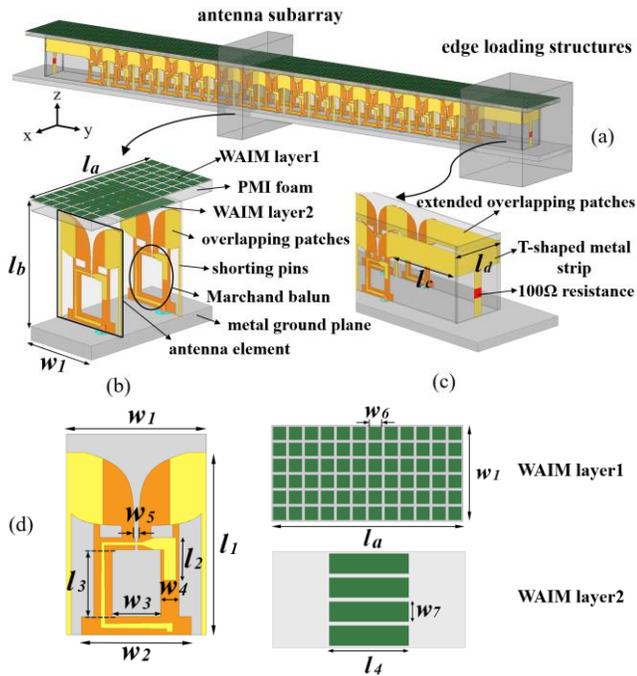

Fig. 1. (a) Geometries of the proposed 2×16 antenna array. (b) Geometries of the proposed antenna subarray. (c) Geometries of the proposed edge loading structures. (d) the detail structure of WAMI layers, balun, dipole, and their corresponding parameters. $l_a$=46, $l_b$=36, $l_c$=36, $l_d$=24, $l_1$=30, $l_2$=7, $l_3$=11, $l_4$=19, $w_1$=23, $w_2$=18, $w_3$=8, $w_4$=3, $w_5$=1, $w_6$=2.9, $w_7$=3.5. (in mm)

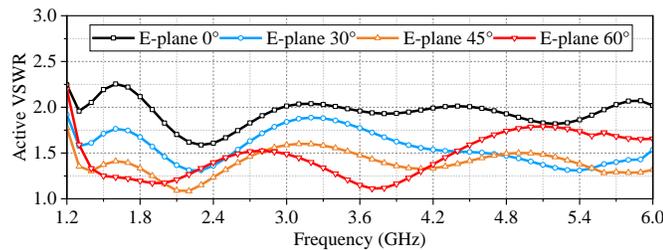

Fig. 2. Simulated active VSWRs of the element for infinite array when steering the main beam to 0°, 30°, 45° and 60° in the E-plane.

conversion from a 50 Ω coaxial connector to a slot line with high characteristic impedance, a printed dipole which is connected to the slot line for excitation, a pair of capacitive overlapping patches and a metal ground plane. The balun, dipole and overlapping patches are designed on both sides of a 0.5mm vertical F4BM substrate ($\varepsilon_{r1}$= 2.2). The overlapping patches are connected to the metal ground plane through shorting pins to suppress the common-mode resonance [15]. The WAIM layer is designed on a 0.127mm horizontal F4BM substrate ($\varepsilon_{r1}$ = 2.2) and a 3mm PMI foam ($\varepsilon_{r2}$ = 1.04) is filled between the two WAIM layers for fixing purpose.

To suppress the ETE, the overlapping patches at the array edges are extended outwards by $l_c$=36mm. Besides, a new type of T-shaped metal strip which is printed on a F4BM substrate ($\varepsilon_{r1}$= 2.2) with a thickness of 0.5mm, is loaded at the end of the extended overlapping patches. The T-shaped metal strip is connected to the extended overlapping patches on the upper side, and is connected to the metal ground on the lower side. A 100Ω resistor is loaded at the bottom of the T-shaped metal strip.

*B. Results and Analysis*

The proposed antenna subarray is simulated in a 1-D

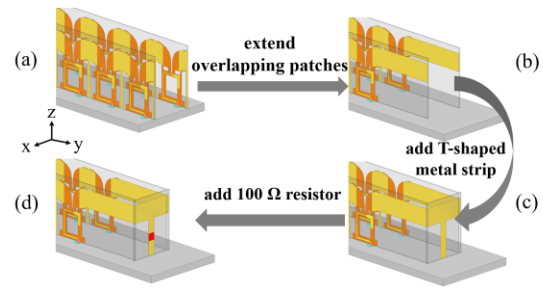

Fig. 3. Design process of the proposed edge loading structures. (a) Model I. (b) Model II. (c) Model III. (d) Model IV.

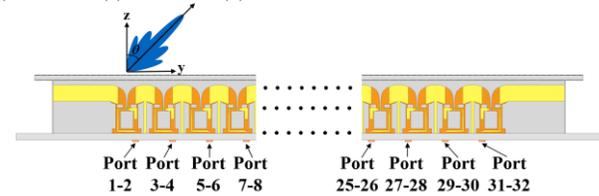

Fig. 4. Port numbering diagram and excitation phase of the proposed 2×16 antenna array.

periodic environment along the *y*-axis. In order to achieve good array radiation performances, the antenna subarray is carefully optimized and designed under periodic conditions. The calculated active VSWRs of antenna element are illustrated in Fig. 2, which shows that the antenna element can feature an operating frequency bandwidth of 1.2~6 GHz for an active VSWR less than 2.3 when scanning up to ±60° in the E-plane. This indicates that if the ETE of the finite array can be suppressed, the proposed TCDLA can realized large scanning angle in the ultrawideband operating band.

Based on the antenna subarray above, a 2×16 array is constructed accordingly. To investigate the effectiveness of the proposed solution, four finite array models with different edge-loading strategies have been involved, as shown in Fig. 3, Model I: 2×16 array without special treatment as a reference, Model II: 2×16 array with extended overlapping patches, Model III: 2×16 array with extended overlapping patches and T-shaped metal strips, Model IV: 2×16 array with extended overlapping patches, T-shaped metal strips and 100 Ω resistors. For convenience in subsequent comparisons, the antenna elements are numbered as shown in Fig. 4, and the angle between the antenna beam direction and the z-axis is *θ*.

To analyze the influence of the ETE on active VSWR, full-wave simulations are performed on the array Model I. The calculated active VSWRs are illustrated in Fig. 5. Since the ETE has more significant impacts at low frequencies, the simulation results are taken from 1.2 GHz to 3 GHz for ease of comparison. Obviously, most peripheral elements at edges or close to edges exhibit high active VSWR (greater than 4) for Model I, especially at low frequencies (1.2~1.6 GHz) of the operating band. The active VSWR deterioration of the Element 1 is the most obvious, which basically stays above 5 within the bandwidth of 1.2~4 GHz. Thus, it is necessary to take some measures to suppress the ETE to improve

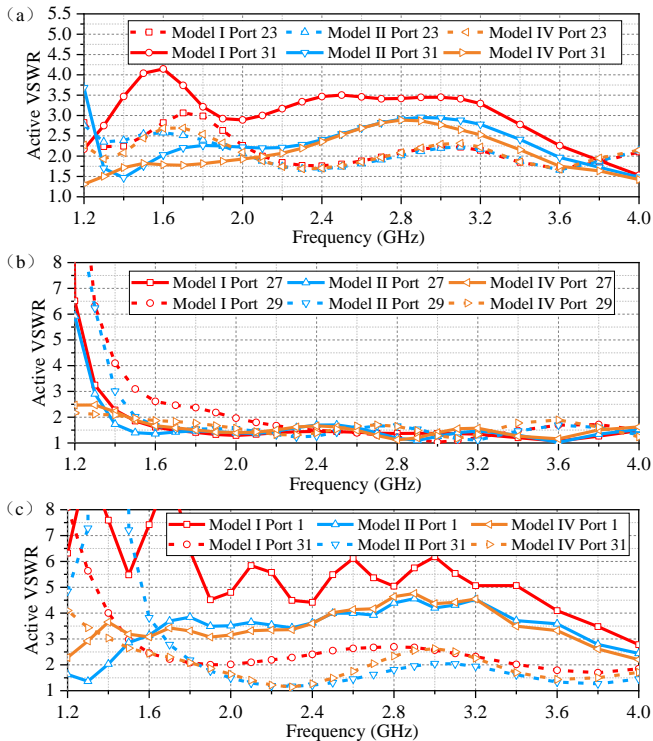

Fig. 5. Active VSWR of edge elements of Models I, II and IV when steering the main beam to (a) 0° and (b)-(c) 60° angles in the E-plane.

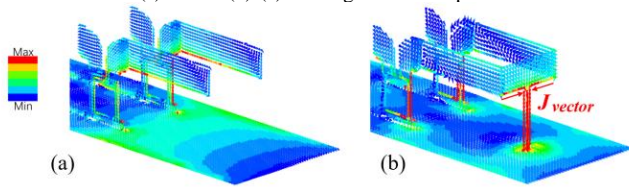

Fig. 6. Vector current distributions at the array edge of (a) Model II and (b) Model III.

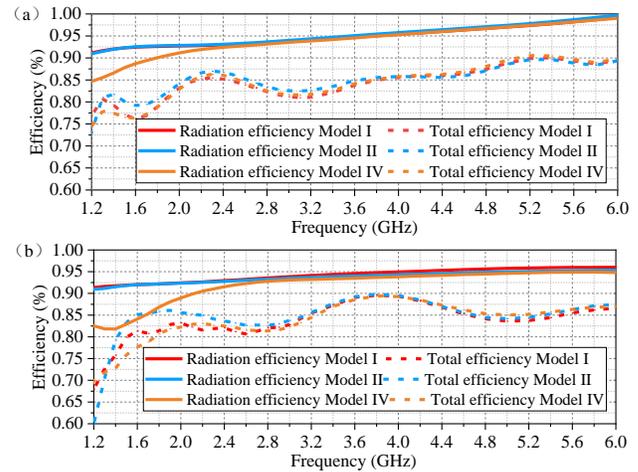

Fig. 7. Radiation efficiency and total efficiency of Models I, II and IV when steering the main beam to (a) 0° and (b) 60° in the E-plane.

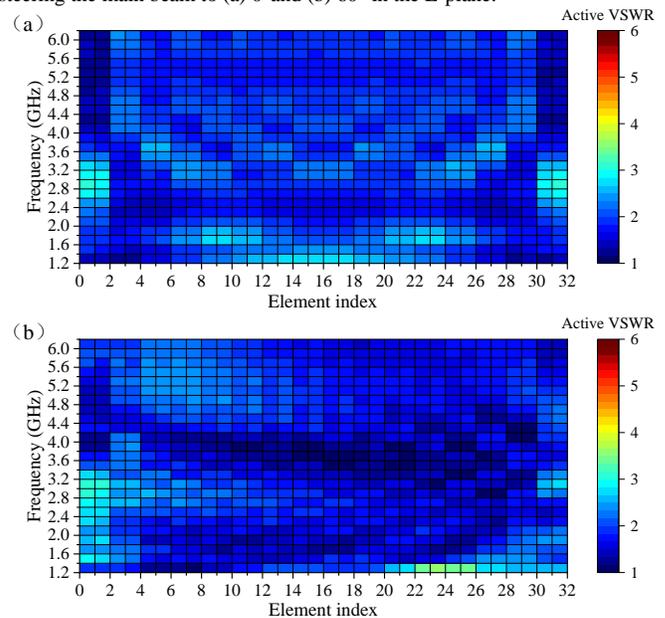

Fig. 8. Simulated active VSWR of each element across the entire operating band when steering the main beam to (a) 0° and (b) 60° in the E-plane.

the performance of the finite array.

In this work, the overlapping patches at the array edges are firstly extended to form a traveling wave structure to guide electromagnetic waves to radiate outward. Furthermore, a new type of T-shaped metal strip which is connected to the extended overlapping patches on the upper side and to the metal ground on the lower side, is added to the array edges. Fig. 6 shows the vector current distributions at the array edge of Models II and III. The loading of the T-shaped metal strip extends the path of current on the extended overlapping patches, which can improve the low-frequency active VSWR of the array edge elements. Besides, it is noted that the currents on the T-shaped structure exhibit equal amplitude and opposite phase along the $x$-axis, thus it will not generate larger cross-polarized radiation. In addition, a 100Ω resistor is soldered at the bottom of the T-shaped metal strip to absorb the left energy. The simulation results in Fig. 5 indicate that after loading the edge structures mentioned above, the active VSWRs of the peripheral elements significantly are reduced to less than 3 except for Port 1 when scanning to 60°. The radiation efficiency and total efficiency of Models I, II and IV are illustrated in Fig. 7, which depicts the radiation efficiency of Models IV can maintain above 82% within the entire operating frequency band.

The excitation phase difference between the edge elements and their neighboring elements is another critical factor contributing to the active VSWR degradation in the edge elements during scanning conditions. Further optimization of the impedance matching for edge elements is achieved by sreducing the excitation phase difference between the edge elements and their neighboring elements. Notably, based on the deterioration trend of the active VSWR of Ports 1 and 31 for Model IV in Fig. 5(c), the operating frequency band is divided into three segments, i.e., 1.2 ~ 2.4 GHz (active VSWR <3.6), 2.4 ~ 4.0 GHz (active VSWR >3.6) and 4.0 ~ 6.0 GHz. Each frequency band employs distinct excitation phase. The optimized excitation phase for the 2×16 linear array is shown in Fig. 4, in which $α = -βdsin(θ)$, $α$ is the excitation phase of the element, $β$ is the wavenumber in free-space, $d$ is the distance between the two neighboring elements, and $θ$ is the beam scanning angle. The maps of the simulated active VSWR for each element across the entire operating band are, respectively, shown in Fig. 8 when the main beam is scanned to 0° and 60° in the E-plane. As can be seen in the figures, the active VSWRs of all array elements successfully decrease to less than 3.5. The operating frequency bandwidth of the 2×16 array has reached that of the antenna element under periodic conditions.

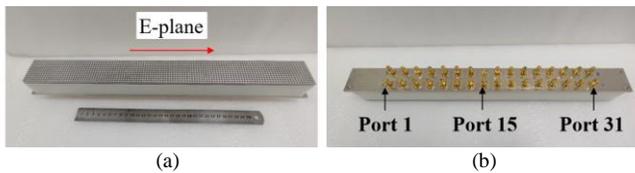

Fig. 9. Photographs of the fabricated 2 × 16 array prototype. (a) Front side (b) Back side.

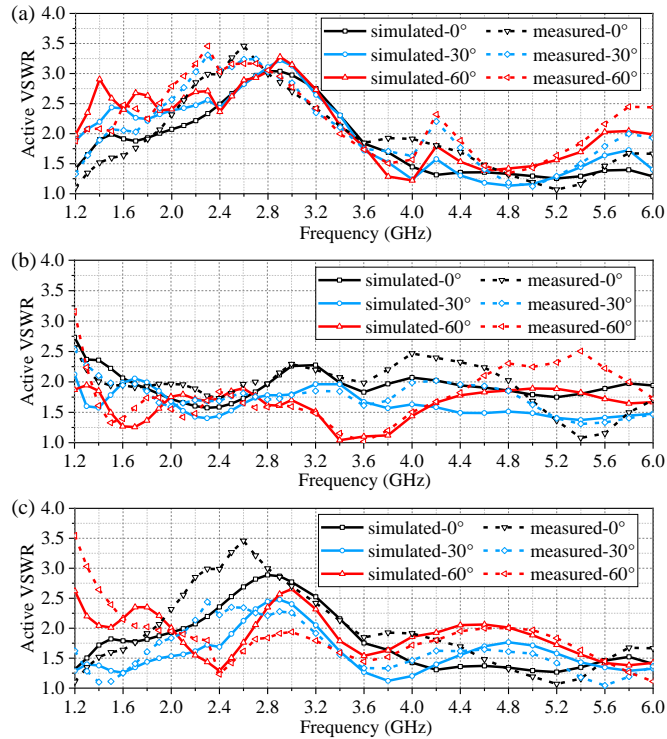

Fig. 10. Simulated and measured active VSWR of edge and central elements for the 2 × 16 array prototype when steering the main beam to 0°, 30° and 60° angles in the E-plane. (a) Port 1. (b) Port 15. (c) Port 31.

TABLE I
COMPARISONS BETWEEN THE PROPOSED ANTENNA AND OTHER WORKS

| Ref. | Approaches | Array type | Impedance Bandwidth (GHz) | Scan range (E/H) | VSWR peak |
|---|---|---|---|---|---|
| [10] | Loading short elements | Planar | 0.2-0.6 (100%) | N.A./±30° | 3 |
| [11] | Amplitude-phase optimization | Planar | 0.2-2 (164%) | ±45°/±45° | 3 |
| [16] | Extended dipole arm | Linear | 2.2-6.0 (86.4%) | ±45°/N.A. | 3 |
| This work | T-shaped strip loaded resistor | Linear | 1.2-6 (133%) | ±60°/N.A. | 3.5 |

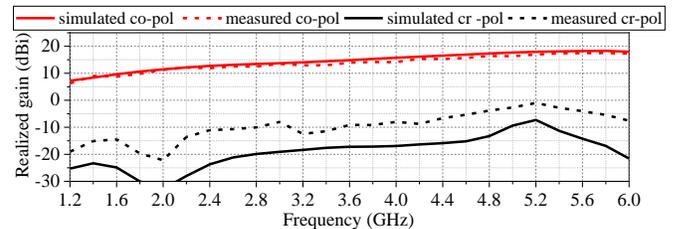

Fig. 11. Co-polarized and cross-polarized gains versus frequency at broadside.

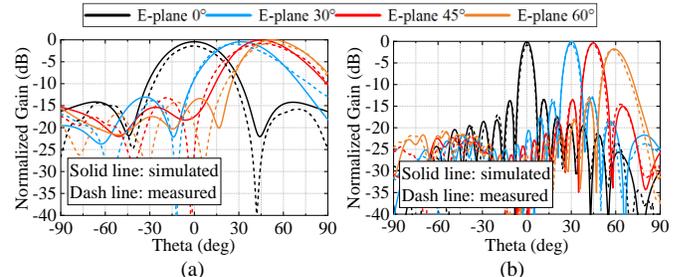

Fig. 12. Simulated and measured co-polarized normalized gain patterns of the 2 × 16 array prototype at (a) 1.2 GHz and (b) 6 GHz when steering the main beam to 0°, 30°, 45° and 60° angles in the E-plane.

The main performance characteristics of the referenced array and this design are shown in Table I. Although the bandwidth of the proposed design is smaller than that in [11], larger scanning angles can be achieved without reducing the maximum power capacity and aperture efficiency of the array. Furthermore, the proposed design demonstrates an enhanced operating bandwidth and scanning angle range compared to the findings in [10] and [16].

III. ARRAY PROTOTYPE MEASUREMENT

A 2×16 prototype has been manufactured and measured, as shown in Fig. 9. The measured active VSWRs for both center and edge elements at broadside and scanning conditions are illustrated in Fig. 10, which shows the measured active VSWRs for both center and edge elements are less than 3.5 under scanning conditions in the E-plane. Fig. 11 shows the measured co-polarized (co-pol) and cross-polarized (cr-pol) gains at broadside. The average difference between the measured co-polarized gain and the simulated values is less than 1 dB across the entire operating bandwidth. Although the measured cross-polarized shows deterioration, the measured cross-polarization ratio across the entire operating bandwidth is below -15 dB. Fig. 12 shows the normalized gain patterns during scanning in E-plane at 1.2 GHz and 6 GHz. As observed, the measured gain patterns match well with the simulated ones. There are some small discrepancies between measurement and simulation results, mainly due to assembly errors.

IV. CONCLUSION

A novel method for mitigating the ETE of finite-size TCDLAs is proposed. Based on extending the length of the overlapping patches, a new type of T-shaped metal strip with a 100 Ω resistor is added to the array edges. Besides, the excitation phase of the elements at the array edges is optimized. The 2×16 finite TCDLA using the proposed method can feature an operating frequency bandwidth of 5:1 (1.2~6 GHz) with an active VSWR<3.5 while scanning up to ±60° in the E-plane. The simulation and measured results effectively demonstrated the effectiveness of the proposed solution in suppressing the ultrawideband ETE of TCDLAs.